\documentclass{article}
\usepackage{spconf,amsmath,graphicx}
\usepackage{enumitem}
\setlist[itemize]{noitemsep, topsep=0pt}
\title{Self-supervised models of speech infer \\universal articulatory kinematics}
\name{Cheol Jun Cho$^{1}$\qquad Abdelrahman Mohamed$^{2}$\qquad \textit{Alan W Black} $^{3}$\qquad  \textit{Gopala K. Anumanchipalli}$^1$}

\address{$^1$UC Berkeley \qquad $^2$ Rembrand \qquad $^3$ Carnegie Mellon University}
\begin{document}

\ninept
\maketitle
\begin{abstract}
Self-Supervised Learning (SSL) based models of speech have shown remarkable performance on a range of downstream tasks. These state-of-the-art models have remained blackboxes, but many recent studies have begun ``probing" models like HuBERT, to correlate their internal representations to different aspects of speech. In this paper, we show ``inference of articulatory kinematics" as fundamental property of SSL models, i.e., the ability of these models to transform acoustics into the causal articulatory dynamics underlying the speech signal. We also show that this abstraction is largely overlapping across the language of the data used to train the model, with preference to the language with similar phonological system. Furthermore, we show that with simple affine transformations, Acoustic-to-Articulatory inversion (AAI) is transferrable across speakers, even across genders, languages, and dialects, showing the generalizability of this property. Together, these results shed new light on the internals of SSL models that are critical to their superior performance, and open up new avenues into language-agnostic universal models for speech engineering, that are interpretable and grounded in speech science.

\end{abstract}
\section{Introduction}
\label{sec:intro}
Self-supervised learning (SSL) has revolutionized every field of machine learning, by providing rich features of natural data without human-annotated labels. Likewise, speech SSL models have been proven to be successful in various speech downstream tasks \cite{yang2021superb}. To understand such utility, the internal representation of speech SSL models has been scrutinized by probing analyses for known speech and linguistic features, such as low-level acoustics, phonetics, and lexical semantics \cite{Pasad2021,Shah2021, pasad2023comparative,cho2023emaprobing}. A comparative analysis by Cho et al. \cite{cho2023emaprobing} demonstrates that the state-of-the-art SSL models are highly correlated with articulatory kinematics and the correlation score can indicate the success of the SSL model in downstream tasks. This finding is extended to developing a high-performance Acoustic-to-Articulatory inversion (AAI) model \cite{wu2023aai}.

Here, we test an intriguing hypothesis -- speech SSL models infer the causal articulatory processes that generate the speech acoustic signal. If this hypothesis is true, such inference should be agnostic to language or dialect, as speech acoustics are a result of the resonance associated with vocal tract shapes that add spectral detail to the vocal source, i.e., vibration of the vocal folds. Besides, the human species has a common canonical structure of vocal tract anatomy and orofacial muscles, regardless of ethnic group, language, and dialect. 

We validate this hypothesis using two approaches: cross-lingual articulatory probing and cross-speaker transferability analysis. First, we probe articulatory representation in speech SSL models trained on different languages, using a comprehensive set of public electromagnetic articulography (EMA) datasets that cover up to 62 speakers from different language and dialect groups. Then, we test how well each individual articulatory system can be transferred to another system, by fitting an affine transformation. Combining these two analyses elucidates the universality of the articulatory representation in speech SSL models. 

While EMA has been a major interface of studying articulatory phonology \cite{browman1992articulatory, ramanarayanan2013spatio, gesture, gan}, the data collection procedure is highly complicated with many physical restrictions that have posed significant barriers in collecting a large scale dataset \cite{rebernik2021review}. Moreover, a majority of the publicly available datasets are concentrated in a few common languages like English.\footnote{This is a significant bottleneck given there are more than 7,000 languages around the world.} Therefore, the development of a universal AAI model can reduce the burden of collecting new EMA data for investigating diverse languages. This is especially valuable for languages with low resources or near extinction where only a short amount of audio is accessible. Our results suggest that the articulatory inference in speech SSL models is transferable across languages, making a critical step toward the universal AAI model. 

Our major findings are:
\begin{itemize}
    \item Articulatory kinematics of different languages and dialects can be recovered from speech SSL models with a simple linear projection, regardless of the language for which the SSL models are trained.
    \item Individual articulatory systems can be aligned by affine transformations, implying the existence of a canonical basis of articulatory kinematics.
    \item While the overall scores are surprisingly high, we observe a preference for language or dialect, providing evidence of language-specific articulatory phonology reflected in speech SSL models.  
\end{itemize}

%\section{Related work}
%\textbf{Speech SSL model}\\
%\textbf{Probing}\\
%\textbf{EMA}\\

\label{sec:intro}

\section{Related Work}
\textbf{Speech SSL.} Wav2vec-2.0 (W2V2) \cite{baevski2020wav2vec} and HuBERT (HB) \cite{hsu2021hubert} have been the most successful SSL models of speech. Both models are trained by masked prediction objectives, where the Transformer encoder predicts randomly masked parts of input features. As raw audio waves are inadequate for target reference, instead, the models predict self-driven units, either from vector quantization (W2V2) or online clustering (H2B) of internal features.\\
\textbf{Cross-lingual speech model.} XLS-R \cite{babu2021xls} and MMS \cite{pratap2023mms} are direct extensions of W2V2, which aim to learn cross-lingual speech representation by training on multiple languages. XLS-R is trained on 50K hours of speech from 53 languages, and MMS is trained on 55K hours of speech from more than 1,000 languages. As the models are trained across languages, they may learn articulatory representation that is shared across languages. \\
\textbf{Probing SSL representation.} Several efforts have been made to explain which features are encoded by SSL. Pasad et al. \cite{Pasad2021, pasad2023comparative} compare lexical semantics with speech SSL models across layers, revealing that lexical semantics rise in the later layer of SSL models. Shah et al. \cite{Shah2021} train non-linear models to probe audio, fluency, pronunciation, and text features, providing a detailed description of information processing in SSL models. Furthermore, clustering analysis suggests that the primary information encoded in SSL models is fine-grained subphonemic units \cite{hubert_unit_1, hubert_unit_2}. Our work is a direct extension of Cho et al. \cite{cho2023emaprobing} that demonstrates evidence of high-fidelity articulatory kinematics in SSL models.\\
\textbf{AAI.} Inversion models often suffer from individual differences in EMA data. A major source of such variability is anatomical differences across speakers. Furthermore, though key articulators are standardized, there is some level of arbitrariness in the locations of the sensors. Together, such misalignment across subjects has challenged developing a speaker independent AAI model. The recent work by Wu et al. \cite{wu2023aai} mitigates the problem by utilizing relative distances between articulators, which are proven to be more consistent across speakers than EMA. This is well aligned with our finding that affine transformations can align individual articulatory systems. 

\section{Methods}
\subsection{Speech SSL models trained for different languages}
\label{sec:model}
We use HB and W2V2 with large sizes (300M parameters) to extract speech representations. We retrieved either of those models from Hugginface, trained with a specific language: English, Mandarin, Korean, French, or Dutch. Furthermore, two cross-lingual W2V2 models are included: XLS-R \cite{babu2021xls} with 300M parameters trained on 53 languages, and MSS \cite{pratap2023mms} with 1B parameters trained on more than 1000 languages. For baseline reference, a set of acoustic features (filter bank, mel spectrogram, and MFCC) are included as well.

\subsection{Comprehensive set of multi-lingual, multi-dialect EMA}
\label{sec:dataset}
Our comprehensive analyses include five public EMA datasets that cover 62 speakers from three languages: English, Mandarin, and Italian (Table \ref{tab:dataset}). For English speakers, there are three regional dialect groups where speakers are from the UK, USA, or China (Beijing or Shanghai). As the EMA data can be highly noisy due to the known instability in the data collection procedure, we filter out three Mandarin speakers from the original corpus. Note that the speakers validated in the original dataset papers are included in the final dataset. Each dataset has a frame-wise annotation of kinematic traces of six articulators: low incisor (LI), upper lip (UL), lower lip (LL), tongue tip (TT), tongue blade (TB), and tongue dorsum (TD)), on the midsagittal plane: X (front-back) and Y (up-down). Each trace is normalized to be zero mean and unit variance.\footnote{The normalization is done per channel and within the clip.} All of these data were collected while naturally speaking full sentences.

\subsection{Probing SSL models with articulatory kinematics}
\label{sec:emaprobing}
We follow the previous articulatory probing approach \cite{cho2023emaprobing} to measure the correlation between SSL representation and articulatory kinematics. Features from 20 ms audio frames are extracted from each layer of each SSL model and projected to the EMA space by fitting a linear inversion model. One difference from the previous study is that a 6 Hz low-pass filter is applied to remove high-frequency noise. The prediction performance is assessed by calculating the correlation between predictions and the ground truth reference, and the results are averaged over a 5-fold cross-validation. %The linear models are fitted for each individual speaker without any regularization, providing acoustic-to-articulation (AAI) inversion models for each of the speakers.

\subsection{Evaluating transferability between speakers}
\label{sec:trans}
We hypothesize that the linear inversion models obtained from the probing analysis are leveraging the same basis of the articulatory subspace that resides in SSL representation. To test this hypothesis, we devise a transferability metric, which measures the correlation between two articulatory systems. Suppose we have two AAI models, \(f_{A}\) and \(f_{B}\), for different speakers, A and B. As there is a variability in the vocal tract anatomy across individuals, we trained another linear model, \(g_{A\rightarrow B}\), to align the articulatory systems, which is trained with L1 regularization weighted with \(\alpha\) = 0.01 to impose sparsity in coefficients. The transferability from speaker A to speaker B is then measured as the correlation of the prediction by these two aligned inversion systems,  \(\text{corr}(g_{A\rightarrow B} \circ f_{A}, f_{B})\). For this analysis, linear inversion models built upon the 17th layer of XLS-R are used and the data from the target speaker (B) is used for training and testing. We only include speakers with high inversion performance (corr \(\geq 0.8\)), totaling 52 speakers. We denote the A and B speaker pair as the source and target pair.

\begin{table}[t]

\caption{Summary of the public EMA datasets used in the analyses. Different language-dialect groups are denoted as EN.\{UK, US, BJ, SH\}: English speakers from the UK; the US; Beijing, or Shanghai, China, respectively, MAN:
Mandarin speakers, and IT: Italian speakers. The number of speakers used for the transferability analysis is denoted in (\(\cdot\)) if different. The gender distribution is balanced in MOCHA-TIMIT, HPRC, and EMA-MAE. }
\vskip 0.0in
\label{tab:dataset}
\begin{center}
\begin{tabular}{c|c|c|c}
\hline
Corpus            & Lang.                                                         & Spk \#                                                            & m/spk                                                \\ \hline
MNGU0 \cite{mngu}       & EN.UK                                                         & 1                                                             & 75                                                   \\ \hline
MOCHA-TIMIT \cite{mochatimit} & EN.UK                                                         & 7                                                             & 27                                                   \\ \hline
HPRC \cite{hprc}        & EN.US                                                         & 8                                                             & 59                                                   \\ \hline
EMA-MAE \cite{emamae}  & \begin{tabular}[c]{@{}c@{}}EN.US\\ EN.BJ\\ EN.SH\end{tabular} & \begin{tabular}[c]{@{}c@{}}20 (18)\\ 10 (9)\\ 9 (5)\end{tabular} & \begin{tabular}[c]{@{}c@{}}12\\ 17\\ 16\end{tabular} \\ \hline
DKU-JNU-EMA \cite{dkujnuema} & MAN                                                           & 4 (2)                                                          & 20                                                   \\ \hline
MSPKA \cite{mspka} & IT                                                            & 3 (2)                                                          & 47                                                   \\ \hline
Total &      --                                                       & 62 (52)                                                          & --                                                   \\ \hline
\end{tabular}
\end{center}
\vspace{-0.3cm}
\end{table}

\section{Results}
\begin{figure*}[t]
\begin{center}
\includegraphics[width=\textwidth,keepaspectratio]{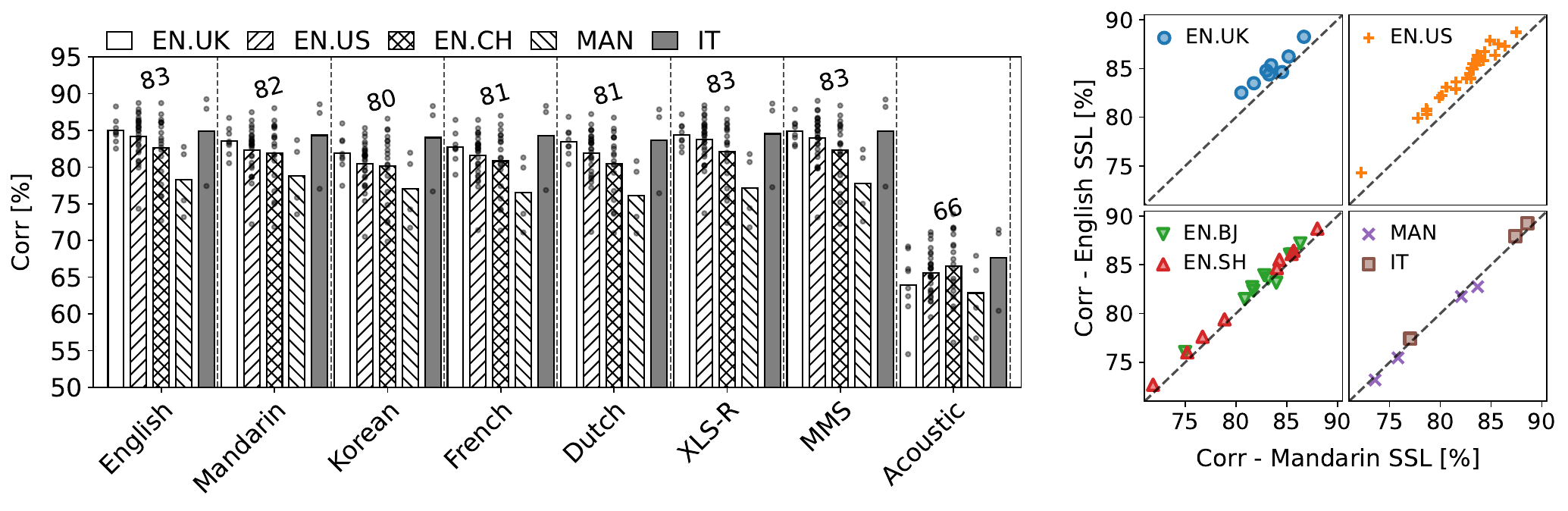}
\vspace{-0.7cm}
  \caption{(\textit{left}) EMA prediction performance of probing SSL models from different languages. The correlations are averaged across 12 EMA channels. Each language-dialect group is denoted by a hatch pattern and each dot denotes an individual speaker. Regardless of the language, the average performance reaches over 0.8. (\textit{right}) Performance comparison of English SSL versus Mandarin SSL, each panel shows specific language-dialect groups. The diagonal dashed lines denote identity lines. For English, native speakers (EN.UK/US) prefer English models over Mandarin models, scattered slightly above the diagonals, but speakers from China (EN.BJ/SH) show almost identical scores.}
  \label{fig:lang_model}
\end{center}
\vspace{-0.3cm}
\end{figure*}
\subsection{Articulatory probing shows high correlations agnostic to training language of SSL models}

Fig.\ref{fig:lang_model} shows the prediction performance of probing SSL models in \S\ref{sec:model} that are trained in different languages. For specific SSL models, the performance of the layer with the highest correlation is reported.\footnote{The layer-wise patterns are omitted due to the limited space, but the patterns are consistent with \cite{cho2023emaprobing}.} For some feature sets with multiple models, the models with higher scores are selected.\footnote{English and Mandarin have both HB and W2V2.} Despite the substantial discrepancy between languages, the average correlations are surprisingly high, reaching over 0.8 regardless of the training language (Fig.\ref{fig:lang_model} \textit{left}), which are far beyond the baseline acoustic features (corr = 0.66\(\pm\)0.040). This indicates that the SSL models naturally learn physical relationships between raw speech signals and vocal tract articulations, which are not bound to specific language, but rather universal properties of the human vocal tract system. Amongst languages, the highest performance is achievable by English SSL models (corr = 0.835\(\pm\)0.039) and cross-lingual models (XLS-R: 0.830\(\pm\)0.039, MMS: 0.832\(\pm\)0.040). The performance difference between MMS and the other two best models is not significant. Given the scale of MMS in both training data and model size, this finding suggests that scaling up has a minimal effect, and learning articulatory representation is readily saturated in the SSL regime.

Probing models for individual speakers show variable performances due to the noisy nature of the data collection, which is mostly induced by the unstable attachment of sensors. Still, 44\% of speakers show high performance above 0.85 correlations, and the best subject achieves 0.89 correlation.\footnote{One of the Italian speakers using the 10th layer of HB trained on English (among 24 layers).} This is remarkable given that we are applying a very simple linear regression on the frozen SSL models that are never exposed to EMA data while training. 

\subsection{Preference for language in articulatory probing}
\label{sec:pref}
Despite the overall high correlations, some noteworthy variability across languages is observable. English models and Mandarin models show higher correlations than models of other languages. As most of the speakers are from either the US, UK, or China, this may indicate some language specificity. However, this difference can not be fully attributed to the language difference because the probing performance is also correlated with in general representational quality of SSL model \cite{cho2023emaprobing}. 

To further test language specificity in SSL models, we compare English SSL models and Mandarin SSL models for each dialect group in English speakers (Fig.\ref{fig:lang_model} \textit{right}). The native English speakers (EN.UK/US) prefer English models over Mandarin models (Fig.\ref{fig:lang_model} \textit{right}; top two panels), to a small extent yet statistically significant (mean diff = 0.019; \(p\) \(<\) \(1e-5\)). However, the English speakers from China (EN.BJ/SH) show almost identical probing performances of these models (Fig.\ref{fig:lang_model} \textit{right}; bottom left panel), showing non-significant difference (mean diff = -0.002; \(p\) = 0.29). As the language is controlled to be English, the observed preference is attributed to differences in articulatory phonology rather than other linguistic components (e.g., semantics or syntax). This indicates that there exist some residual articulatory habits originating from speaking Mandarin as a mother tongue language. The other language groups, MAN and IT, show relatively minor differences and the number of speakers is too few to evaluate the significance.

\subsection{Individual articulatory systems are mutually transferable with affine transformations}

\begin{figure*}[t]
\begin{center}
\includegraphics[width=\textwidth,keepaspectratio]{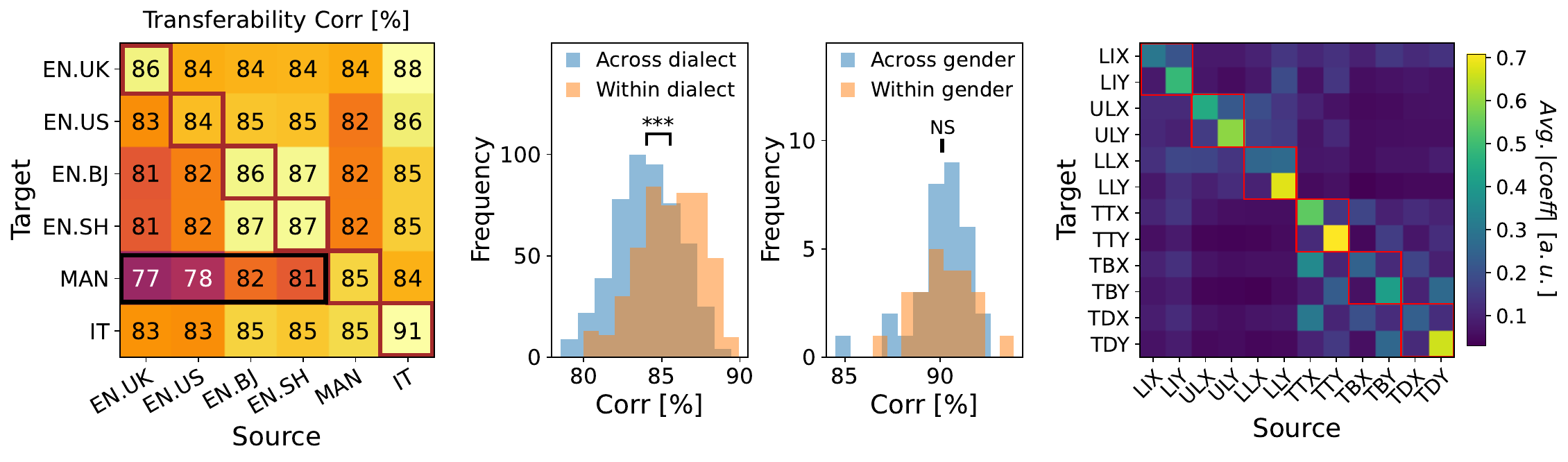}
\vspace{-0.7cm}
  \caption{(\textit{far-left}) Correlation matrix denoting transferabilities between language-dialect groups. The scores are averaged over possible pairs between groups. (\textit{mid-left}) Distributions of correlations across dialects (blue) and within dialects (orange). The English speakers from China (EN.SH+EN.BJ) and the native English speakers from the US (EN.US) groups in the EMA-MAE dataset are used. (\textit{mid-right}) Distributions of correlations across gender (blue) and within gender (orange). Four male and four female speakers from the HPRC dataset are used. (\textit{far-right}) The average absolute coefficients in the affine transformations between the articulatory systems.}
  \label{fig:trans_results}
  \vspace{-0.3cm}
\end{center}
\end{figure*}

For each pair of speakers, we measure the transferability scores (\S\ref{sec:trans}) between the articulatory systems obtained from the probing analyses. The far-left matrix of Fig.\ref{fig:trans_results} shows the average correlations between source and target group pairs. The scores within groups are strikingly high, showing near or above 0.85 average correlation (diagonal brown boxes). This suggests that the articulatory systems of individual speakers are affine transformations of each other. Such high correlations are also observable in the transformations across language-dialect groups, especially between the Italian (IT) and the native English speakers (EN.UK/US), which show even higher scores than within-group scores. The IT has in general high transferability to others, indicating that the AAI models built upon one of these particular speakers can serve as the closest universal AAI model. %This is well depicted in an example (Fig.\ref{fig:pred}). 
These high-fidelity affine transformations between different speakers demonstrate a significant level of isometry in the human articulatory system. This is strong evidence that speech SSL models infer the canonical basis of the articulatory kinematics. 

\begin{figure}[t]
\begin{center}
\centerline{\includegraphics[width=180pt]{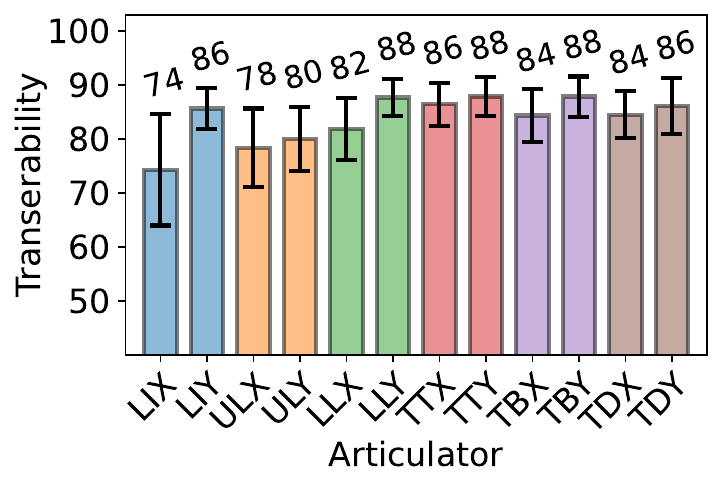}}
\vspace{-0.3cm}
\caption{Transferability scores of each articulator averaged over all source-target pairs.}
\label{fig:art}
\vspace{-0.7cm}
\end{center}
\end{figure}

The coefficients in the affine transformations are mostly assigned to the same articulators (Fig.\ref{fig:trans_results} \textit{far-right}; red boxes). This suggests that the transformations are geometric alignments that correct the variability induced by the difference in vocal tract anatomy or the sensor locations. While most of the six articulators show definite self-assignments, the tongue blade (TB) is affected by other tongue parts, which is natural since the tongue is highly deformable and TB bridges the tongue tip and dorsum. Fig.\ref{fig:art} shows transferability scores for each articulator. Some channels, LIX, LIX, ULY, and LLX, are relatively difficult to align, which reflects their innate subtlety in shaping phonetic information. %For example, the front and back movements of the jaw, LIX, are relatively subtle compared to the movements along the Y axis. %This pattern across articulators is also reflected in the example (Fig.\ref{fig:pred}). 

\subsection{The variance in transferability reflects language-specific articulatory phonology}
\label{sec:trans_group}
When we compare the within and across dialects transferability, the across dialect transformations show significantly lower scores than those of the within dialect cases (Fig.\ref{fig:trans_results} \textit{mid-left}). However, when groups are divided by gender, there is no significance between the within and across gender transformations (Fig.\ref{fig:trans_results} \textit{mid-right}). As gender is the most significant factor of anatomical differences, this result suggests that the transferability between articulatory systems is not affected by the vocal tract anatomy. Therefore, the observed patterns across languages and dialects are more likely due to differences in articulatory phonology. This is consistent with the language preference in cross-lingual probing \S\ref{sec:pref}. To avoid variance induced by the data collection site, we confined analyses on dialect groups and gender groups to the EMA-MAE corpus and HPRC corpus, respectively.

This language-specific pattern is also reflected in the transferability matrix (Fig.\ref{fig:trans_results} \textit{far-left}; black box). While the transferability scores from the English groups (EN.*) to the Mandarin group (MAN) are generally lower than in other cases, the Chinese English groups (EN.BJ/SH) are more transferable than the native English speakers (EN.UK/US). Such a pattern is not observed in the opposite cases (MAN \(\rightarrow\) EN.*).

%\begin{figure}[t]
%\begin{center}
%\centerline{\includegraphics[width=200pt]{sample_temp.png}}
%\caption{Example predictions by a target speaker's AAI model (TRG, blue, EN.UK), and by an aligned source speaker's AAI model (SRC, orange, IT). The black lines (GT) are ground truth signals.  }
%\label{fig:pred}
%\end{center}
%\end{figure}

\section{Discussion}

We demonstrate that the recent speech SSL models can recover articulatory kinematics with simple linear mapping, achieving high performance inversion regardless of speakers, languages, and dialects. This is further verified by finding affine transformations from one articulatory system to another. Our findings provide strong evidence that there is a canonical basis of articulatory phonology which is naturally emerging in self-supervised learning of speech. 

Our findings evoke another interesting hypothesis that articulatory kinematics are not only the physical interface of speech but also the continuous embedding representations of phonetics. As the SSL models are trained by the masked prediction objective without any external labels, the resulting representations are perceptual descriptions of how sounds are shaped in natural speech data, which also shows high correspondence to the human auditory perception \cite{li2022dissecting}. Here, we demonstrate that the natural selection of the perceptual embedding space of speech is a continuous, dynamic articulatory space. Particularly, the group-level analyses in \S\ref{sec:trans_group} reveal the language specificity in articulatory systems but little evidence of gender specificity that entails large anatomical differences. This observation further supports the hypothesis of equivalence between articulatory kinematics and continuous phonetic embeddings.

The articulatory information captured by EMA is not complete as limited to the midsagittal plane, and upper bounded by the inevitable noise ceiling of the sensors. Also, the source information such as vocal fold vibration is missing. However, our findings indicate that a complete set of articulatory kinematics may lie in SSL representations which are missing in EMA. This is an interesting future work toward unsupervised discovery of articulatory kinematics without relying on EMA.

%Speech SSL transforms one-dimensional waveform to the continuous embedding space of phonetic information. And the articulatory kinematics lie on a linear subspace of such continuous space. Therefore, the articulatory kinematics can be recovered with high-fidelity with a simple linear mapping from such space. Moreover, the continuous phonetic embedding space largely overlaps across languages as we found that the high-perofrmance of linear inversion can be achieved agnostic to language or dialect. However, there is some non-negligible preference over languages which indicates a variance in articulatory phonology across languages. 
%TODO:// point out the best subject case and its average transferability. And how that can serve as a universal articulatory inversion model.

%TODO:// potential explanation for errors: EMA is not a complete space of articulatory phonology + sensing procedure itself is noisy. 

%TODO:// SSL models are learned without label and just learned to represent internal statistics of speech. And this suggest that the embedding of such intenrnal statistics is articulatory kinematics.

%TODO:// potential hypothesis in fundamental theory of phonology, that the continuoos embedding space of phonetics is articulatory kinematics. 

\section{Acknowledgements}

This research is supported by the following grants to PI Anumanchipalli --- NSF award 2106928, BAIR Commons-Meta AI Research, the Rose Hills Innovator Program, and UC Noyce Initiative, at UC Berkeley. Special thanks to Shang-Wen (Daniel) Li for discussions and for providing advice.
\vfill\pagebreak
\bibliographystyle{IEEEbib}
%\fontsize{10}{11}\selectfont
\bibliography{refs}
\label{sec:refs}
\end{document}